\documentclass[preprint,showpacs,preprintnumbers,amsmath,amssymb]{revtex4}

\usepackage{graphicx}
\usepackage{dcolumn}
\usepackage{bm}

\begin{document}


\title{Strange dibaryon resonance in the $\bar{K}NN$-$\pi Y N$ system}

\author{Y. Ikeda and T. Sato}
 \email{ikeda@kern.phys.sci.osaka-u.ac.jp, tsato@phys.sci.osaka-u.ac.jp}
\affiliation{%
Department of Physics, Graduate School of Science,
Osaka University, Toyonaka, Osaka 560-0043, Japan}%

\date{\today}

\begin{abstract}
Three-body resonances in the $\bar{K}NN$ system have been studied within
a framework of the $\bar{K}NN-\pi Y N $ coupled-channel Faddeev equation.
By solving the three-body equation, the energy dependence of
the resonant $\bar{K}N$ amplitude is fully taken into account.
The S-matrix pole has been investigated 
from the eigenvalue of the kernel
with the analytic continuation of the  scattering amplitude
on the unphysical Riemann sheet.
The $\bar KN$ interaction is constructed from
the leading order term of the chiral Lagrangian using relativistic
kinematics.
The $\Lambda(1405)$ resonance is dynamically
generated in this model,
where the $\bar{K}N$ interaction parameters are
fitted to the data of scattering length.
As a result we find a three-body resonance of 
the strange dibaryon system with 
binding energy, $B \sim 79$ MeV, and 
width, $\Gamma \sim 74$ MeV.
The energy of the three-body resonance is found to be
sensitive to the model of the $I=0$ $\bar KN$ interaction.
\end{abstract}

\pacs{
11.30.Rd, 11.80.Jy, 13.75.Jz }
\maketitle

\section{Introduction}
\label{intro}

The analysis of the  kaonic atoms~\cite{Gal1} 
revealed an attractive  $\bar{K}$-nucleus interaction.
Although the strength of the attraction depends on the parametrization
of the density dependence of the optical potential~\cite{Gal1}
and the theoretical study of the $\bar{K}$ optical potential
suggests a rather shallow potential~\cite{Ram}, there has been a great
interest in the possibilities of $\bar{K}$-nucleus bound states 
in recent years.
Akaishi and Yamazaki~\cite{Aka1,Aka2} studied
the kaon bound states in light nuclei and found
deeply bound kaonic states, for example, $B \sim 100$ MeV for $^3_{\bar{K}}$H.
In their study, the kaonic nuclear states were 
investigated by using the $\bar{K}$ optical potential,
which is constructed by folding the 
g matrix with a trial nuclear density.
The potential model of  $\bar{K}N-\pi\Sigma$ interaction
is determined to reproduce the $\Lambda(1405)$ and the scattering length.
The kaonic nuclear states are further studied by using
a method of  antisymmetrised molecular dynamics~\cite{Dote} using the
$\bar{K}N$ g matrix.

Among the simplest $\bar{K}$ nucleus state,
the $K^-pp$ state, which has 
strangeness $S=-1$, total angular momentum and parity $J^\pi=0^-$, 
and isospin  $I=1/2$ dibaryon state,
is expected to have largest component of the $I=0$ $\bar{K}N$.
An experimental signal of the $K^-pp$ bound state
is reported by the FINUDA Collaboration
from the analysis of the invariant mass distribution
of $\Lambda-p$ in the $K^-$ absorption reaction on nuclei~\cite{Agne}.
The reported central value of the binding energy, $B$, and the width, $\Gamma$,
are $(B,\Gamma)=(115,67)$ MeV, which is below the $\pi \Sigma N$ threshold
energy. This data may be compared with the
predicted values $(B,\Gamma) = (48,61)$ MeV in Ref.~\cite{Aka2}.
However it was pointed out that the data can be understood by
the two-nucleon absorption of $K^-$ in nuclei together with the 
final state interaction of the outgoing baryons~\cite{Magas}.

In the attractive interaction of kaon in nuclei,
the resonance $\Lambda(1405)$  in the  s-wave and $I=0$ channel
$\bar{K}N$ scattering state plays an essential role. 
The energy of the $\Lambda(1405)$ is 
below the $\bar{K}N$ threshold and strongly couples with
the $\pi\Sigma$ state.  
Although the kaonic nuclear states
have been studied so far by using the $\bar{K}N$ g matrix or
optical potential, it might be very important to 
examine the full dynamical calculation of $\bar{K}N-\pi\Sigma$
system by taking into account the
energy dependence of the resonance t matrix and
the coupling with the $\bar{K}N-\pi\Sigma$ channel explicitly.
Such a theoretical study may be possible in the
simplest kaonic nuclei with baryon number $B=2$ system.
In this work, we study the
strange dibaryon system by taking into account the three-body
dynamics using the $\bar{K}NN-\pi\Sigma N-\pi\Lambda N$ 
$(\bar{K}NN-\pi YN)$ coupled-channel Faddeev
equation with relativistic and non-relativistic kinematics.

Methods to investigate resonances in the three-body system
have been developed in the studies of the
three-neutron~\cite{Glock,Moll},
$\pi NN$ dibaryon~\cite{Afnan2,Matsu} and $\Sigma NN$ hypernuclei~\cite{Matsu,Pear,Afnan}.
In this work, we employ a method 
started by Gl\"ockle~\cite{Glock} and M\"oller~\cite{Moll}
and developed by  Matsuyama and Yazaki, Afnan, 
Pearce and Gibson~\cite{Matsu,Pear,Afnan}
to find a pole of the S matrix in the unphysical energy plane
from the eigenvalue of the kernel of the Faddeev equation.
To analytically continue the scattering amplitude into the
unphysical sheet, the path of the momentum integral must be
carefully deformed in the complex plane to avoid
possible singularities.

The most important  interaction for the study of the
strange dibaryon system is for the $I=0$ $\bar{K}N$ states.
The internal structure of the $\Lambda(1405)$ has been a long standing issue.
The chiral Lagrangian~\cite{Oller,Jido,Bor} approach can 
describe well the low energy $\bar{K}N$ reaction with the 
meson-baryon dynamics.
A genuine $q^3$ picture of the $\Lambda(1405)$ 
coupled with meson-baryon~\cite{Hama} 
may not yet be excluded.
Though previous studies of the $\bar{K}NN$ system
used phenomenological models of the $\bar{K}N$ potentials,
we use  s-wave meson-baryon coupled-channel potentials
guided by the lowest order chiral Lagrangian.
With this model, the strength of the potentials
and the relative strength of the potentials among various 
meson-baryon channels are not parameters but are determined
from the SU(3) structure of the chiral Lagrangian. 
In this model, the $\Lambda(1405)$ is an 'unstable bound state', whose
pole on the unphysical sheet will become the bound state of $\bar{K}N$
when the coupling between the $\bar{K}N$ 
and the $\pi Y$ is turned off.
We examine a relativistic model as well as a nonrelativistic model
to account for the relativistic energy of pion in the  $\pi Y N$ state.

We briefly explain our $\bar{K}NN-\pi Y N$ coupled-channel equations 
and the procedure to search for the three-body resonance in section 2.
The model of the two-body 
interactions used in this work is explained in section 3.
We then report our results on the $\bar{K}NN$ dibaryon resonance in section 4.
This work is the extension of the early version of our analysis
reported in Ref.~\cite{ikeda}.
Recently Shevchenko {\it et al}.~\cite{she} performed a similar study of the
$\bar{K}NN$ system using Faddeev equation starting form the
phenomenological $\bar{K}N$ interaction within a nonrelativistic
framework. The comparison of our results with theirs will be discussed
in section 4.

\section{Coupled channel Faddeev equation and resonance pole}
We start from the Alt-Grassberger-Sandhas(AGS) equation~\cite{ags} 
for the three-body scattering problem.
The operators $U_{i,j}$ of the three-body scattering  satisfy the 
AGS equation
\begin{eqnarray}
U_{i,j} & = & (1 - \delta_{i,j})G_0^{-1} +
\sum _{n\neq i}t_n G_0U_{n,j}.\  \label{Eqags-1}
\end{eqnarray}
Here we label the pair of particles $j,k$ by
the spectator particle $i=1,2,3$.
The two-body  t matrix $t_i$  of particles $j,k$ with the spectator
particle $i$ is given by the solution of the Lippmann-Schwinger equation:
\begin{eqnarray}
t_i & = & v_i + v_i G_0 t_i.
\end{eqnarray}
Here $G_0 = 1/(W - H_0 + i\epsilon)$ is the free Green's function 
of the three particles, and $W$ is the total energy of the three-body system.

When the two-body interactions $v_i$ are given in separable form
with the vertex form factor $|g_i>$ and the coupling constant $\gamma_i$ as
\begin{eqnarray}
v_i & = & |g_i>\gamma_i < g_i|,
\end{eqnarray}
the AGS-equation of Eq. (\ref{Eqags-1}) is written in the form
\begin{eqnarray}
X_{i,j}(\vec{p}_i,\vec{p}_j,W) & = & 
(1-\delta_{i,j})Z_{i,j}(\vec{p}_i,\vec{p}_j,W) + 
\sum _{n\neq i} \int d \vec{p}_n 
    Z_{i,n}(\vec{p}_i,\vec{p}_n,W)
   \tau _n (W)
X_{n,j}(\vec{p}_n,\vec{p}_j,W).\nonumber \\
 \label{Eqags-2}
\end{eqnarray}
The amplitude $X_{i,j}$ is defined by  the matrix element of 
$U_{i,j}$ between state vectors $G_0|\vec{p}_i,g_i>$ as
\begin{eqnarray}
X_{i,j}(\vec{p}_i,\vec{p}_j,W) & = & <\vec{p}_i,g_i|G_0U_{i,j}G_0|\vec{p}_j,g_j>.
\end{eqnarray}
The state vector $|\vec{p}_i,g_i>$ represents
a plane wave state of the spectator $i$ and 
the state vector $|g_i>$ of the interacting pair.

\begin{figure*}
\resizebox{0.8\textwidth}{!}{%
\includegraphics{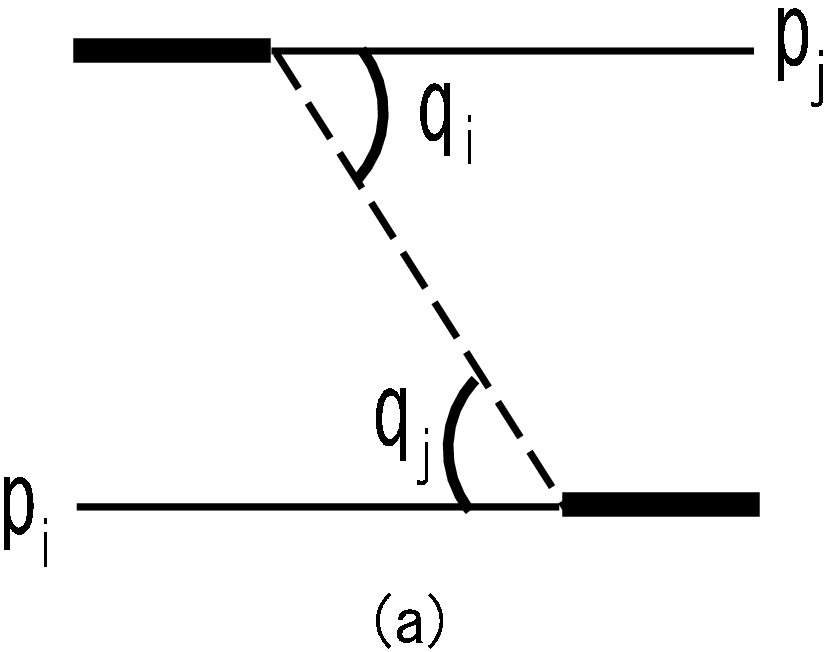}\hspace*{2cm}
\includegraphics{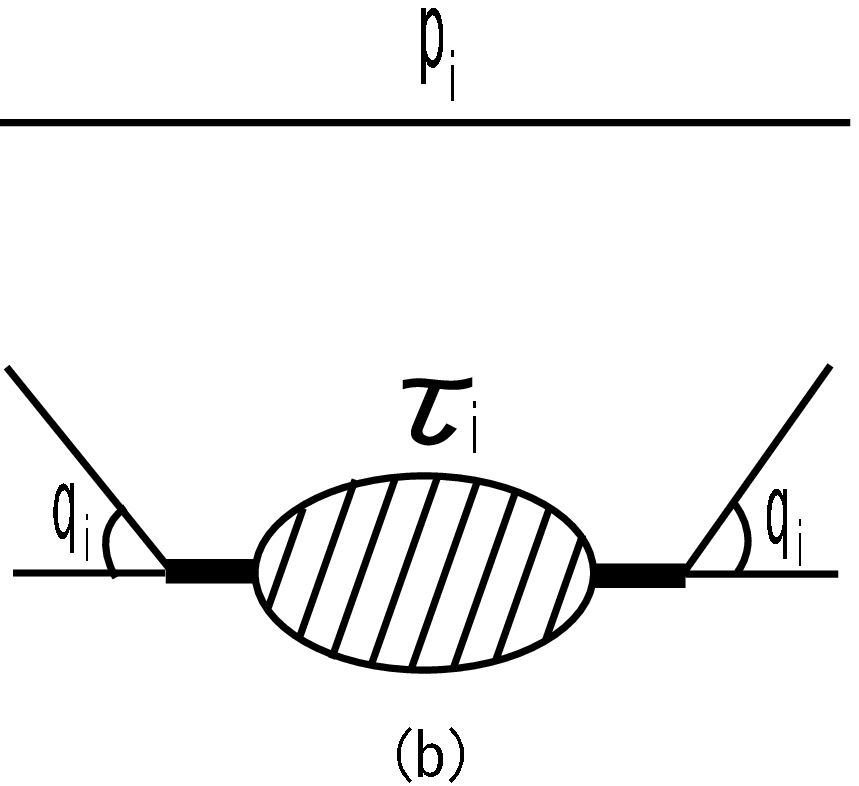}}
\caption{
Graphical representation of
(a) one particle exchange interaction
$Z_{i,j}(\vec p_i,\vec p_j,W)$ 
and (b) two-body t-matrix  $\tau _i(W)$.
The relative momentum of the interacting particles
is  given by $\vec{q}_i$ for spectator particle $i$.
}
\label{diag-Ztau}
\end{figure*}

The driving term $Z_{i,j}$ of Eq. (\ref{Eqags-2})
shown in Fig. \ref{diag-Ztau}(a)
is given by the particle exchange mechanism defined as
\begin{eqnarray}
Z_{i,j}(\vec{p}_i,\vec{p}_j,W) & = & <\vec{p}_i,g_i|G_0|\vec{p}_j,g_j>  \\
  & = &
\frac{g^*(\vec{q}_i) g(\vec{q}_j)}
{W- E_i(\vec{p}_i) -E_j(\vec{p}_j) -E_{k}(\vec{p}_k)}. \label{Z}
\end{eqnarray}
Here the momentum of the exchanged particle $k (\neq i,j)$
is given as $\vec{p}_k = - \vec{p}_i - \vec{p}_j$ and
$g(\vec{q}_i)$ is the vertex form factor of the
two-body interaction $g(\vec{q}_i)= <g_i|\vec{q}_i>$.
The energy $E_i(\vec{p}_i)$ is given by
$E_i(\vec{p}_i) = m_i + \vec{p}_i^2/2m_i$ for the nonrelativistic
model and $E_{i}(\vec{p}_i) =\sqrt{m_i^2 + \vec{p}_i^2}$ for the
relativistic model.
The relative momentum is given by
$\vec{q}_i = (m_k\vec{p}_j - m_j\vec{p}_k)/(m_j + m_k)$
for the nonrelativistic model, while
we define $q_i = |\vec{q}_i|$ for the relativistic model as
\begin{eqnarray}
q_i & = &
\sqrt{(\frac{W_i^2+m_j^2-m_k^2}{2W_i})^2-m_j^2},\\
W_i & = &
\sqrt{(E_j(\vec p_j)+E_k(\vec p_k))^2-\vec p_i^2}.
\end{eqnarray}

The two-body t matrix can be solved for the separable interaction as
\begin{eqnarray}
t_i & = & |g_i> \tau _i(W) <g_i|.
\end{eqnarray}
Here the 'isobar' propagator $\tau_i$, illustrated in
Fig. \ref{diag-Ztau}(b), is given as
\begin{eqnarray}
\tau_i(W) & = & [1/\gamma_i - \int d\vec{q}_i 
\frac{  |g_i(\vec{q}_i)|^2}
{W - E_i(\vec{p}_i) - E_{jk}(\vec{p}_i,\vec{q}_i)}]^{-1}.
\end{eqnarray}
The two-body t matrix depends on the energy $E_i(\vec{p}_i)$ of the
spectator particle.
Here $E_{jk}$ is the energy of the interacting pair given
as $E_{jk}(\vec{p}_i,\vec{q}_i)= m_j + m_k + \vec{p}_i^2/(m_j+m_k) + 
\vec{q}_i^2/\mu_i$ for the non-relativistic model, while
 $E_{jk}(\vec{p}_i,\vec{q}_i)=
\sqrt{(E_j(\vec{q}_i)+E_k(\vec{q}_i))^2  + \vec{p}_i^2}$
for the relativistic model.
The reduced mass is defined as $\mu_i = m_jm_k/(m_j+m_k)$.

Following the standard method of angular momentum expansion~\cite{book},
the AGS equation reduces to following
coupled integral equations by keeping only s-wave states:
\begin{eqnarray}
X_{i,j}(p_i,p_j,W) & = & Z_{i,j}(p_i,p_j,W) + \sum_n \int dp_n p_n^2 
\nonumber \\ &\times&
 K_{i,n}(p_i,p_n,W) X_{n,j}(p_n,p_j,W). \label{Eqags-3}
\end{eqnarray}
Here we used a simplified notation for the kernel $K=Z\tau$,
which can be written as
\begin{eqnarray}
K_{i,n}(p_i,p_n,W) = 2 \pi \int d(\hat{p}_i \cdot \hat{p}_n)
\frac{g^*(q_i)g(q_n) }
{W-E_i(p_i)-E_j(p_n)-E_k(\vec{p}_i+\vec{p}_n)}\tau _n(W). \label{Eqags-4}
\end{eqnarray}

The formulas given above are valid for the spinless and distinguishable
particles without channel coupling among the Fock-space vectors.
In our  $\bar{K}NN$ resonance problem, we have included
the following $\bar{K}NN$ and $\pi Y N$ states:
\begin{eqnarray}
| a > & = &  | N_1, N_2, \bar{K}_3>,\\
| b > & = &  | N_1, Y_2, \pi _3>,\\
| c > & = &  | Y_1, N_2, \pi _3> ,
\end{eqnarray}
with $Y_i$ is  $\Sigma_i$ or $\Lambda _i$.
After anti-symmetrizing the amplitude for identical particles
of nucleons~\cite{Afnan2},
we obtain the following forms of the coupled AGS equations,
\begin{eqnarray}
&&
\left(
\begin{array}{c}
X_{Y_K,Y_K} \\
X_{Y_{\pi},Y_K} \\
X_{d,Y_K} \\
X_{N^*,Y_K}
\end{array} 
\right)
=
\left(
\begin{array}{c}
Z_{Y_K,Y_K} \\
0 \\
Z_{d,Y_K} \\
0
\end{array} 
\right)
\nonumber \\
&&
+
\left(
\begin{array}{cccc}
-Z_{Y_K,Y_K}\tau _{Y_K,Y_K} &
-Z_{Y_K,Y_K}\tau _{Y_K,Y_{\pi}} &
2Z_{Y_K,d}\tau _{d,d} &
0
\\
0 & 0 & 0 &
-Z_{Y_{\pi},N^*}\tau _{N^*,N^*}
\\
Z_{d,Y_K}\tau _{Y_K,Y_K} &
Z_{d,Y_K}\tau _{Y_K,Y_{\pi}} &
0 & 0
\\
-Z_{N^*,Y_{\pi}}\tau _{Y_{\pi},Y_K} &
-Z_{N^*,Y_{\pi}}\tau _{Y_{\pi},Y_{\pi}} &
0 & 0
\end{array} 
\right) 
\left(
\begin{array}{c}
X_{Y_K,Y_K} \\
X_{Y_{\pi},Y_K} \\
X_{d,Y_K} \\
X_{N^*,Y_K}
\end{array} 
\right). \label{ags-full}
\end{eqnarray} 
Here we have suppressed the spin-isospin quantum numbers, 
the spectator momentum $p_j$ and the total
energy of the three-body system $W$ in $Z$, $X$ and $\tau$ for simplicity.
The concise notation of $Y_K$, $Y_\pi$, $d$ and $N^*$ represents 
the 'isobars' and their decay channels.
The decay channels of isobars $Y_K$, $Y_\pi$, $d$ and $N^*$ are
$\bar{K}N(I=0,1)$, $\pi\Sigma(I=0,1)$  and $\pi\Lambda(I=1)$,
$NN(I=1)$ and $\pi N(I=1/2,3/2)$, respectively.
Here $I$ is isobar isospin.
Those indices uniquely specify the three-body states of $X$ and $Z$
except $N^*$ showing $\Sigma N^*$ and $\Lambda N^*$.
Therefore we have a nine-channel coupled  equation
of Eq. (\ref{ags-full}) for spin singlet, s-wave three-body system.
The explicit form of Eq. (\ref{Z}) when we include spin-isospin
is summarized in the Appendix.

The dominant Fock space component is expected to be
$|\bar{K}NN>$, and therefore 
the most important amplitudes are $X_{Y_K,Y_K}$ and $X_{d,Y_K}$.
They couple to each other through the kaon exchange $Z_{Y_K,Y_K}$
and nucleon exchange $Z_{Y_K,d}$ mechanisms.
Notice, however, the $\pi YN$ component is also implicitly included
in $\tau_{Y_K,Y_K}$ when we solve the two-body $\bar{K}N - \pi Y$ 
coupled-channel equations. 
The $\pi Y N$ components, $X_{Y\pi,Y_K}$ and $X_{N^*,Y_K}$,
couple with the $\bar{K}NN$ components through the pion exchange mechanism
$Z_{N^*,Y_\pi}$ and the $\pi N$ and $\pi Y$ 'isobars' 
$\tau_{N^*,N^*}$, $\tau_{Y_\pi,Y}$. The pion exchange mechanism
may play an important role in the width of the resonance.
In this work, we have not included weak $YN$ interaction.
It was found in Ref.~\cite{she}
that the $YN$ interaction plays rather minor role in
this strange dibaryon system.

To find the resonance energy of the three-body system
using the AGS equation of Eq. (\ref{ags-full}),
we follow the method used in Refs.~\cite{Glock,Moll,Matsu,Pear,Afnan}.
The AGS equation of Eq. (\ref{Eqags-4}) is 
a Fredholm-type integral equation with the kernel $K=Z\tau$.
Using the eigenvalue $\eta_a(W)$ and the eigenfunction $|\phi_a(W)>$
of the kernel for given energy $W$,
\begin{eqnarray}
Z\tau|\phi_a(W)> & = & \eta_a(W)|\phi _a(W)>,
\end{eqnarray}
the scattering amplitude $X$ can be written as
\begin{eqnarray}
X & = & \sum_a \frac{|\phi_a(W)><\phi_a(W)|Z}{1 - \eta_a(W)}.
\end{eqnarray}
At the energy $W=W_p$ where $\eta_a(W_p)=1$, the amplitude
has a pole, and therefore $W_p$ gives the bound state or resonance energy.

\begin{figure*}
\resizebox{0.5\textwidth}{!}{%
\includegraphics{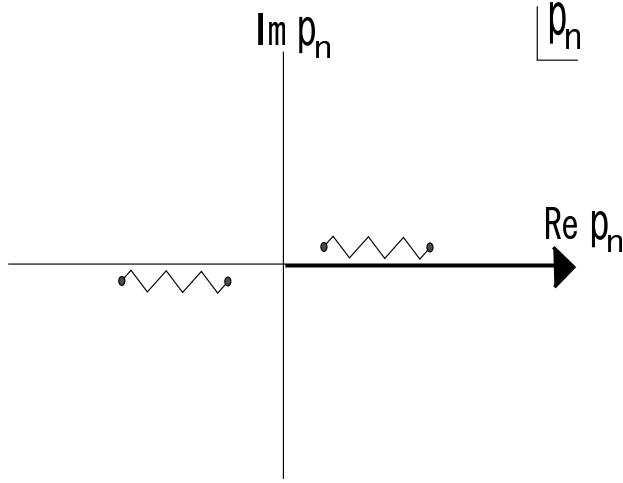}
}
\caption{The singularities of the one particle exchange
interaction $Z(p_i,p_n,W)$ in the complex $p_n$ plane
at $W=E+i \epsilon$ and the real $p_i$.
}
\label{disc-real}
\end{figure*}

Since a resonance pole appears on the unphysical energy Riemann sheet,
we need analytic continuation of the scattering amplitude.
We use here the nonrelativistic model to explain a method
of analytic continuation, which is based on Refs.~\cite{Moll,Matsu}.
At first we examine the singularities of the kernel
of Eq. (\ref{Eqags-4}).
Above the threshold energy of the three-body break up $W > m_i + m_j + m_k$,
$Z(p_i,p_n,W)$ has logarithmic singularities.
The branch points appear at $p_n = \pm p_{Z_{1,2}}$, where
\begin{eqnarray}
p_{Z_1} & = & -\frac{\mu _j}{m_k} p_i
+\sqrt{2 \mu _j W_{th} -\frac{\mu _j}{\eta _j} p_i ^2}, \\
p_{Z_2} & = & +\frac{\mu _j}{m_k} p_i
+\sqrt{2 \mu _j W_{th} -\frac{\mu _j}{\eta _j} p_i ^2},
\end{eqnarray}
with
\begin{eqnarray*}
\mu _j  & = & \frac{m_i m_k}{m_i + m_k},\\
\eta _j & = & \frac{m_j (m_i + m_k)}{m_i + m_j + m_k},\\
W_{th}  & = & W - m_i -m_j - m_k.
\end{eqnarray*}
For given $p_i>0$, the cuts run
from $p_{Z_1}$ to $p_{Z_2}$ above the positive real axis of
complex $p_n$ plane and from $-p_{Z_1}$ to $-p_{Z_2}$
below the negative real axis as shown in Fig. \ref{disc-real},
while the integration of momentum $p_n$ in Eq. (\ref{Eqags-4})
is along the real positive axis.

\begin{figure*}
\resizebox{0.5\textwidth}{!}{%
\includegraphics{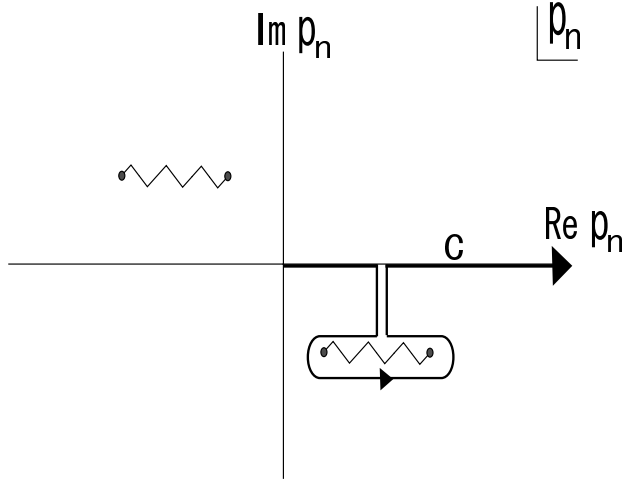}
}
\caption{The integration contour $C$
and the singularity of $Z$ at
 $W=E - i \Gamma /2$ and real value of $p_i$.}
\label{disc-cmp}
\end{figure*}

Let us consider the case when $W$ has a negative imaginary part.
For given $p_i>0$, the cut from $p_{Z_1}$ to $p_{Z_2}$
moves into the fourth quadrant across the
integration contour of $p_n$.
Assuming the integrand of  Eq. (\ref{Eqags-3}) is an analytic function
around real positive $p_n$,
one can perform an analytic continuation of the amplitudes
by deforming the integration contour 
along the logarithmic singularity as shown in Fig. \ref{disc-cmp}
and then we obtain amplitudes on the unphysical Riemann sheet.

\begin{figure*}
\resizebox{0.5\textwidth}{!}{%
\includegraphics{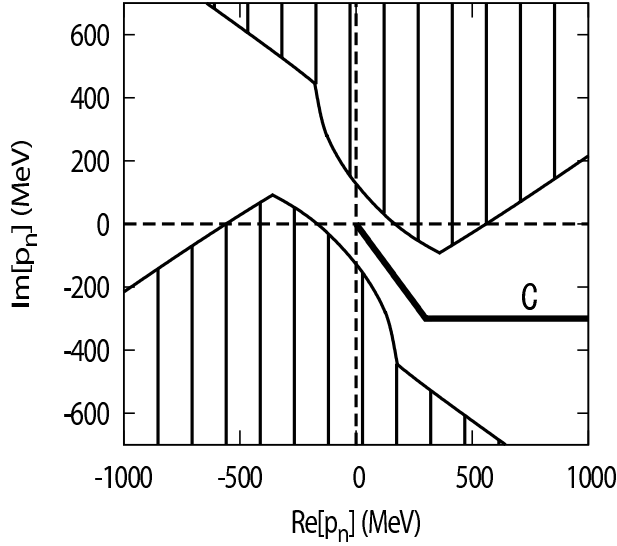}
}
\caption{The logarithmic singularities of 
the $\pi$ exchange mechanism $Z(p_i,p_n,Z)$ at $W_{th}=10-i35$ MeV
in the complex $p_n$ plane.
$C$ is integration  contour of $p_n$ and $p_i$.
}
\label{pi-ex}
\end{figure*}

In principle it might be possible to solve the AGS equation keeping
the momentum variables real and taking into account the
discontinuity across the cut. 
The moving logarithmic singularities depending on
$p_i$ make it difficult to solve the integral equation.
To overcome this problem we deform the integration
contour of $p_i$, $p_n$, into the fourth quadrant of the
complex momentum plane so that we take into account the
contribution of the cuts.
As an example of our $\bar KNN-\pi Y N$ problem,
we choose  the integration contour of $p_n$ as shown in solid line
in Fig. \ref{pi-ex}. Here we take the energy 
$W= 10 - i35 +m_{\pi}+m_{\Sigma}+m_N$ MeV, which is below the mass of $\bar K NN$
and above the $\pi Y N$.
The shaded region in Fig. \ref{pi-ex} shows 
the cuts of 'Z'  for the pion exchange mechanism.
The cuts become 'forbidden regions' because the
position of the cuts depends on $p_i$, which runs the same
integration contour as $p_n$.
In our numerical calculation, we studied all the 'forbidden regions' 
for $\pi, N$ and $K$ exchange mechanism and determined the integration contour.
With the integration contour C in Fig. \ref{pi-ex}, we choose the physical
sheet of $\bar{K}NN$.

\begin{figure*}
\resizebox{1.0\textwidth}{!}{%
\includegraphics{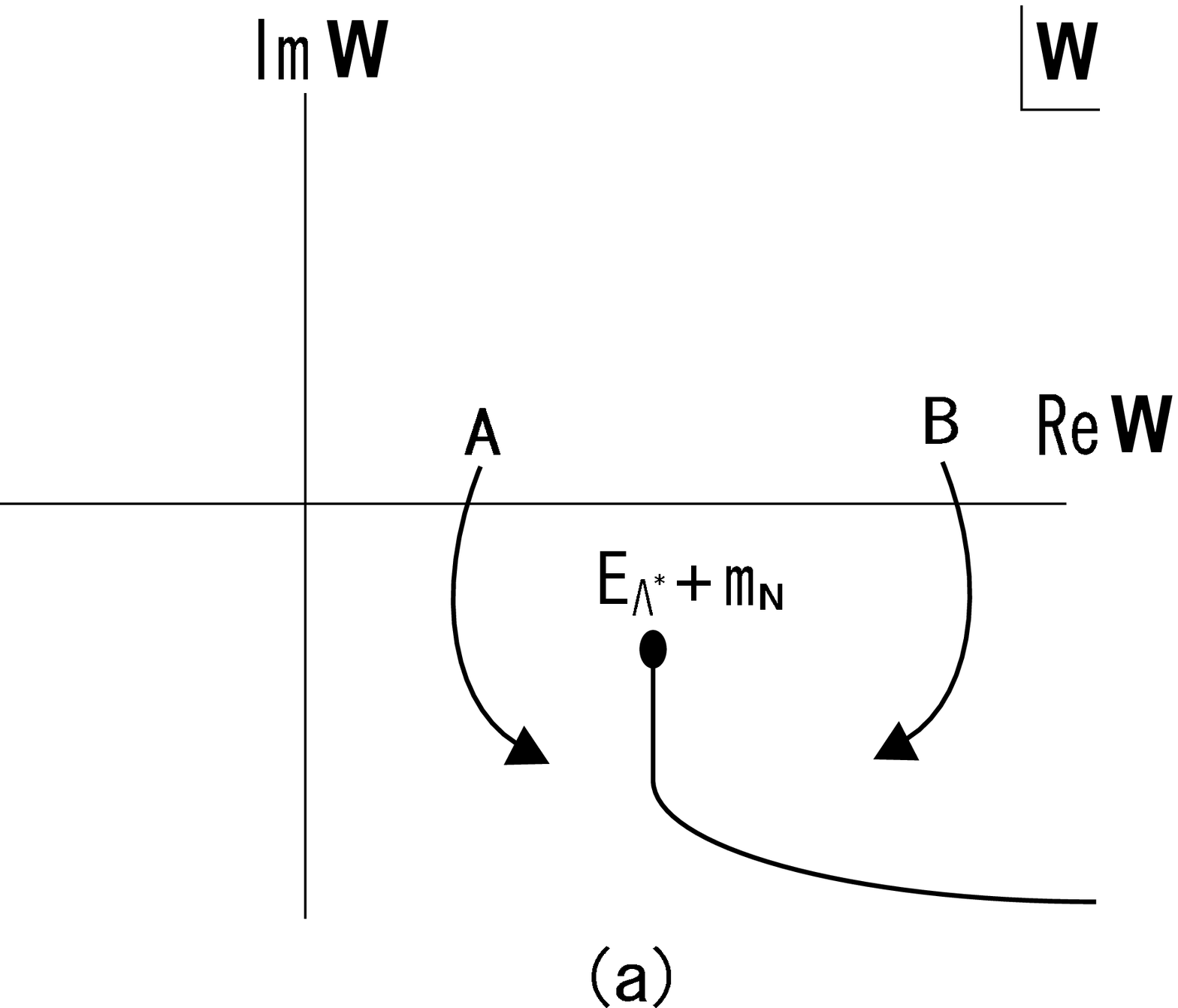}\hspace*{0.5cm}
\includegraphics{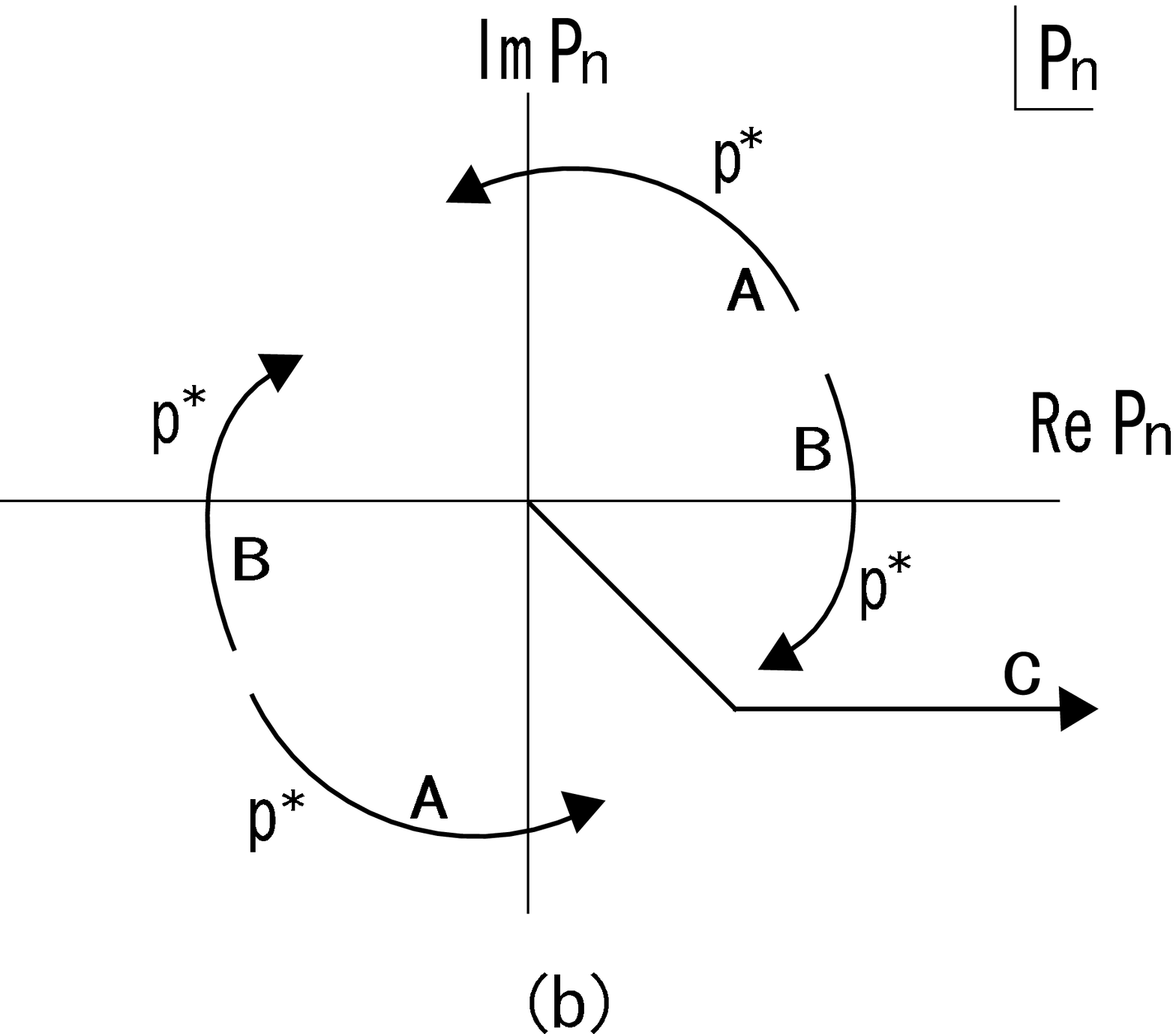}}
\caption{The singularities due to the three-body resonance and 
the $\Lambda(1405)$
in (a) the complex energy plane and (b) the momentum plane.}
\label{lam-cut}
\end{figure*}

The singularities of the isobar propagator $\tau(W)$ 
arises from the three-body Green's function in the 
integrand of $\tau$. The poles are at
$q_n = \pm \sqrt{2 \mu _j W_{th} -\frac{\mu _j}{\eta _j} p_i^2}$.
Since $q_s= (p_{Z_1}+p_{Z_2})/2$, 
we can analytically continue it into the same unphysical sheet as the case in $Z$
as long as
we keep the same deformed contour as the one used in $Z$.
Another singularity we have to worry about is the singularity due to the
two-body resonance.
Since our $\bar{K}N-\pi\Sigma$ system 
has the two-body resonance $\Lambda(1405)$,
the cut starts from the two-body resonance energy in the complex energy
plane.
To examine this, we write the approximate energy dependence of the
$\tau$ as
\begin{eqnarray}
\tau_i(W)
\sim \frac{1}
{W -\frac{p_N ^2}{2 \eta_N} -E_{\Lambda ^*} -m_N}. \label{tau}
\end{eqnarray}
Here $p_N$ and $m_N$ are the momentum and mass of the 
spectator nucleon.
The reduced mass of the spectator nucleon
with the isobar pair $\bar KN$ or $\pi \Sigma$ is denoted as $\eta_N$
and $E_{\Lambda ^*}$ is the pole energy of $\Lambda(1405)$.
At $W =\frac{p_N ^2}{2 \eta_N} +E_{\Lambda ^*} +m_N$ with $p_N$
on the contour C in Figs. \ref{lam-cut} (b),
the two-body t-matrix has a singularity, which is plotted
as a solid line in Fig. \ref{lam-cut}(a).
We illustrate the typical trajectories of the three-body resonance pole  $W= W_p$ 
as curves A and B in Fig. \ref{lam-cut}.
If the pole trajectories A and B intercept the two-body  $N\Lambda(1405)$
cut, then the analytic continuation to the  $N\Lambda(1405)$ unphysical energy
sheet must be examined.  
The same situation from the $p_N$ plane
is shown in Fig. \ref{lam-cut} (b).
The momentum $p^*$ corresponding to the energy $W_p$ of the three-body resonance
is determined by
\begin{eqnarray}
p^* = \pm \sqrt{ 2 \eta_N 
(W_{p}-E_{\Lambda ^*}-m_N) }. \label{plambda}
\end{eqnarray}
If $p^*$ intercepts the contour C, we have to take care of the
analytic continuation of the  $N\Lambda(1405)$ energy sheet.
As will be seen in section 4, the trajectories of the
three-body resonance in our calculation
follow line A of Fig. \ref{lam-cut}(a) and do not intercept the 
singularity of the two-body resonance.

\section{Model of the two-body interactions}
We take into account the $\bar{K}N$ interactions in $J^\pi=1/2^-$, $I=0$ and
$I=1$ states, the $\pi N$ interactions in $J^\pi=1/2^-$,
$I=1/2$ and $3/2$ states and the $NN$ interaction in $I=0,^1S_0$ state.
Our s-wave meson-baryon interaction
is guided by the 
leading order effective chiral Lagrangian
for the octet baryon $\psi_B$ and the pseudoscalar meson $\phi$ fields
given as
\begin{eqnarray}
L_{int} & = & \frac{i}{8F_\pi^2}tr(\bar{\psi}_B \gamma^\mu
                                 [[\phi,\partial_\mu \phi],\psi_B]).
\end{eqnarray}
The meson-baryon potential derived from the chiral Lagrangian can be written as
\begin{eqnarray}
<\vec{p}',\beta|V_{BM}|\vec{p},\alpha>
 & = & - C_{\beta,\alpha}\frac{1}{(2\pi)^3 8F_\pi^2}
\frac{m_\beta + m_\alpha}
{\sqrt{4E_{\beta}(\vec{p}')E_\alpha(\vec{p})}} \nonumber \\ & \times &
g_\beta(\vec{p}')g_\alpha(\vec{p}).
\label{mb-int}
\end{eqnarray}
Here $\vec{p}$ and $\vec{p}'$ are the momentum of the meson in the
initial state $\alpha$ and the final state $\beta$.
The strength of the potential at zero momentum is not an  arbitrary
constant but is determined by the pion decay constant $F_\pi$.
The relative strength between the meson-baryon states 
is controlled by the constants $C_{\beta,\alpha}$ which are basically
determined by the SU(3) flavor structure of the chiral Lagrangian.
The parameter of our model is the cutoff $\Lambda$
of the phenomenologically introduced vertex function 
$g_\alpha(\vec{p})=\Lambda_{\alpha }^4/(\vec{p}^2 + \Lambda_{\alpha }^2)^2$.

The most important interaction for the study of the $\bar{K}NN$ system
is the $I=0$ $\bar{K}N$ interaction. We describe the $\bar{K}N$ interaction by the
coupled-channel model of the $\bar{K}N$ and the $\pi\Sigma$ states.
The constants $C_{\beta,\alpha}$ for this channel are 
given  as $C_{\bar{K}N-\bar{K}N}=6,C_{\bar{K}N-\pi\Sigma}=-\sqrt{6}$
and $C_{\pi\Sigma-\pi\Sigma}=8$.
The cutoff $\Lambda$ is determined by fitting the
scattering length $a^{I=0}_{\bar{K}N}=-1.70 + i0.68$ fm of
Ref.~\cite{Mart}. The values of $\Lambda$ are around
1 GeV and are given as model (a) in tables \ref{hyo3}
and \ref{hyo4} for the nonrelativistic and the relativistic models.
In general, the form factors of the relativistic models are 
hard compared with those of the non-relativistic models because of the
weak relativistic kinetic energy.
We found a resonance pole at $W=1420-i30$ MeV for
the non-relativistic and the relativistic models.
The relativistic kinematics might be important in describing $\pi Y$
channel because of the small pion mass.
We choose this model (a) as a standard parameter of the $\bar{K}N$
interaction. 

The $\bar{K}N$ scattering lengths are not very well constrained 
from the data.
The ranges of the $\bar{K}N$ scattering lengths are studied 
within the chiral unitary model in Ref.~\cite{Bora2}.
In this work, we simply examined models with the scattering length
$a^{I=0}_{\bar{K}N}=(-1.70 \pm 0.10) + i(0.68 \pm 0.10)$ fm
in order to examine the sensitivity of the energy of the three-body resonance 
on the input model of the two-body interaction.
The cutoff $\Lambda$'s for those models are given
as models (b)-(e) of Tables \ref{hyo3} and \ref{hyo4}.
The values of the resonance energy are about $1415 \sim 1425$ MeV,
and the width $50 \sim 70$ MeV,
which are close to the values of the chiral model in Ref.~\cite{Bor}.
One can notice that there is a correlation between
the real (imaginary) part of the pole energy of the $\Lambda(1405)$ 
and the imaginary (real) part of the scattering length.
Those resonance energies are slightly larger than
the pole energy reported in Ref.~\cite{Dlitz}.
Therefore as a last model, model (f) reproduces
the deeper resonance energy $1406-i25$ MeV of Ref.~\cite{Dlitz}.
The scattering length of this model is  $-1.72+i0.44$ fm, 
which is, however, somewhat different from the value $-1.54+i0.74$ fm 
in Ref.~\cite{Dlitz}.

The $I=1$ $\bar{K}N$ interaction is described by the
$\bar{K}N-\pi\Sigma-\pi\Lambda$ coupled-channel model.
The coupling constants $C_{\beta,\alpha}$ are 
$C_{\bar{K}N-\bar{K}N}=2$, 
$C_{\bar{K}N-\pi \Sigma}=-2$,
$C_{\bar{K}N-\pi \Lambda}=-\sqrt{6}$, 
$C_{\pi \Sigma -\pi \Sigma}=4$ and
$C_{\pi \Sigma -\pi \Lambda}=C_{\pi \Lambda -\pi \Lambda}=0$.
The cutoff $\Lambda$'s are determined to fit
the imaginary part of the scattering length of Ref.~\cite{Mart},
which are given as model (A) in Tables \ref{hyo5} and \ref{hyo6}
for the non-relativistic and the relativistic models.
The real part of the scattering length of those models is
larger than $a^{I=1}_{\bar{K}N}= 0.37 + i0.60$ fm of Ref.~\cite{Mart}.
The $K^-p$ scattering length predicted from model (aA)
, which is model (a) for $I=0$ and model (A)
for $I=1$ interactions,  is 
between the central values of the two kaonic hydrogen data~\cite{Iwa,Itoh,Dear}.
To study the sensitivity of the models of $I=1$ $\bar{K}N$ interaction
to the resonance energy of $K^-pp$ system, we constructed model (B)
given in Tables \ref{hyo5} and \ref{hyo6}. 
A similar model of $\bar{K}N$ interaction is developed to
study $K^-d$ scattering\cite{bahaoui}.
The range of the vertex form factor found in in Ref. \cite{bahaoui},
which is monopole form factor with 880MeV cutoff mass,
is comparable to ours.

\begin{figure*}
\resizebox{1.0\textwidth}{!}{%
\includegraphics{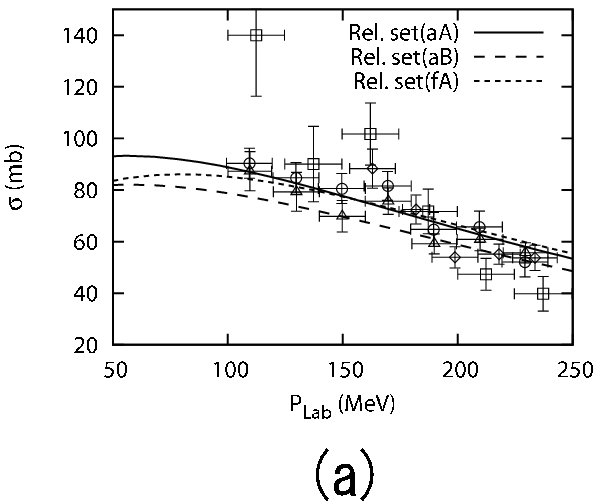}\hspace*{0.3cm}
\includegraphics{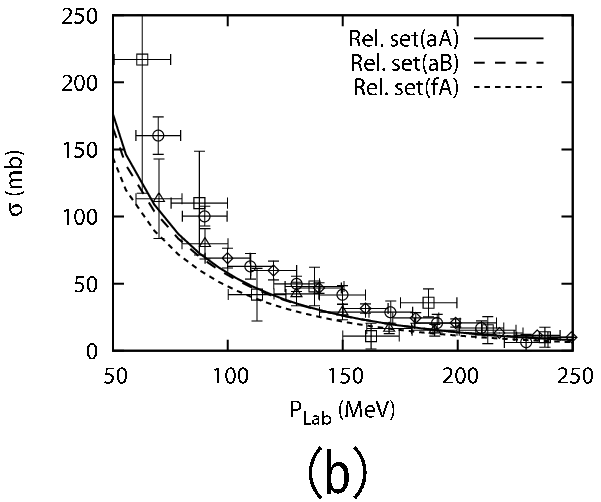}\hspace*{0.3cm}
\includegraphics{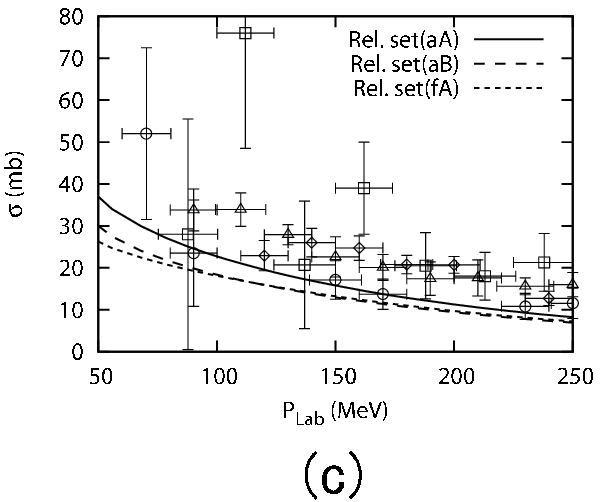}\hspace*{0.3cm}}
\resizebox{0.67\textwidth}{!}{%
\includegraphics{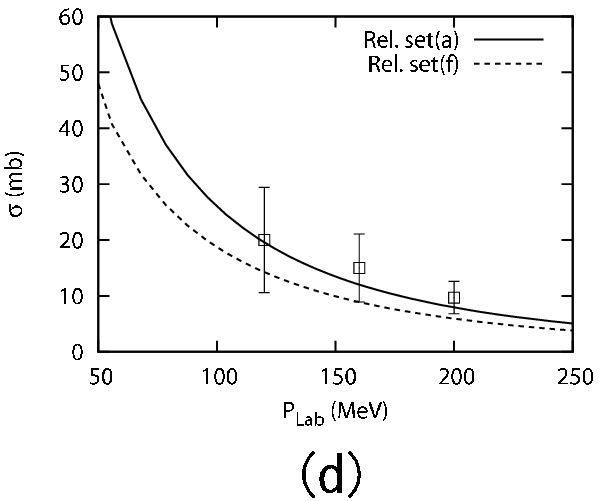}\hspace*{0.3cm}
\includegraphics{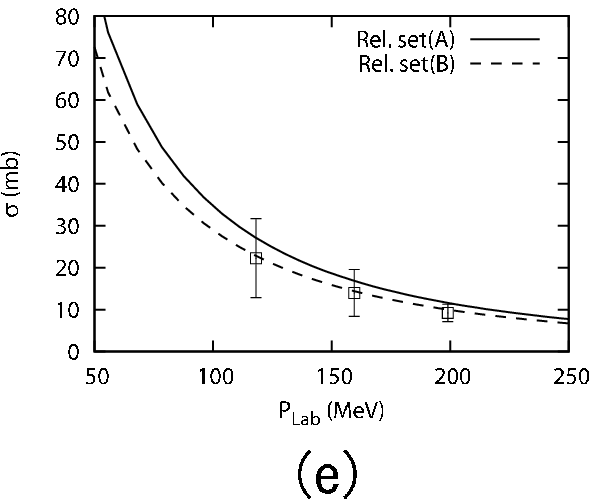}}
\caption{The total cross section of 
(a) $K^-p \rightarrow K^-p$,
(b) $K^-p \rightarrow \pi^+\Sigma^-$
(c) $K^-p \rightarrow \pi^-\Sigma^+$ 
(d) $K^-p \rightarrow \pi^0\Sigma^0$ and
(e) $K^-p \rightarrow \pi^0\Lambda$ reactions
in the relativistic model.
The solid (dashed, dotted) curve shows the result using
model (aA) ((aB), (fA)). Data are taken from 
Ref.~\cite{crs1,crs2,crs3,crs4,crs5}.
}
\label{kp-cros}
\end{figure*}

The total cross sections of $K^- p$ reactions predicted
from our models (aA), (aB) and (fA) are shown in Fig. \ref{kp-cros}
together with the data~\cite{crs1,crs2,crs3,crs4,crs5}.
The models (aA) and (aB)
describe well the $K^-p \rightarrow K^-p$(Fig. \ref{kp-cros}a),
$K^-p \rightarrow \pi^+\Sigma^-$(Fig. \ref{kp-cros}b) and
$K^-p \rightarrow \pi^-\Sigma^+$(Fig. \ref{kp-cros}c) reactions,
where both $I=0$ and $I=1$ interactions contribute to the cross section. 
The models of $I=0$ ($I=1$) can be tested from 
$K^-p \rightarrow \pi^0\Sigma^0$(Fig. \ref{kp-cros}d)
($K^-p \rightarrow \pi^0\Lambda$(Fig. \ref{kp-cros}e)) reactions, where
models 'a' and 'A/B' describe the cross sections well.
The model (fA) tends to give smaller cross sections. 
It is noticed, however, as we will see, that the resonance energy
of the $K^-pp$ system is more sensitive to the $I=0$ $\bar{K}N$
interaction and less sensitive to $I=1$ interactions, while 
both $I=0$ and $I=1$ interactions are equally important to describe
$K^-p$ cross sections and kaonic hydrogen data.

\begin{table*}[htbp]
\begin{center}
\begin{tabular}{c||cc|cc}
   &$\bar{K}N$(MeV)&$\pi\Sigma$(MeV)&
Scattering Length(fm) & Resonance energy(MeV)\\ \hline \hline
(a) & 1095 & 1450 & $-1.70+i0.68$ & $1419.8-i29.4$\\ 
\hline
(b) & 1105 & 1550 & $-1.60+i0.68$ & $1422.2-i33.7$\\ 
\hline
(c) & 1085 & 1350 & $-1.80+i0.68$ & $1418.5-i25.0$\\ 
\hline
(d) & 1120 & 1340 & $-1.70+i0.59$ & $1414.6-i29.4$\\ 
\hline
(e) & 1070 & 1540 & $-1.70+i0.78$ & $1424.3-i28.3$\\ 
\hline
(f) & 1160 & 1100 & $-1.72+i0.44$ & $1405.8-i25.2$\\ 
\hline
\end{tabular}
\caption{The cutoff parameters, scattering length
and the resonance pole of the relativistic models of
I=0 $\bar{K}N - \pi \Sigma$ interaction.}
\label{hyo3}
\end{center} 
\end{table*}

\begin{table*}[htbp]
\begin{center}
\begin{tabular}{c||cc|cc}
   &$\bar{K}N$(MeV)&$\pi\Sigma$(MeV)&
Scattering Length(fm) & Resonance energy(MeV)\\ \hline \hline
(a) & 946 & 988  & $-1.70+i0.68$ & $1420.1-i30.1$\\ 
\hline
(b) & 954 & 1035 & $-1.60+i0.68$ & $1422.4-i34.7$\\ 
\hline
(c) & 940 & 944  & $-1.80+i0.68$ & $1418.7-i26.0$\\ 
\hline
(d) & 968 & 933  & $-1.70+i0.58$ & $1414.3-i30.5$\\ 
\hline
(e) & 927 & 1031 & $-1.70+i0.78$ & $1424.7-i29.0$\\ 
\hline
(f) & 1000 & 800 & $-1.72+i0.43$ & $1404.8-i25.5$\\ 
\hline
\end{tabular}
\caption{The cutoff parameters, scattering length
and the resonance pole of the nonrelativistic models of
I=0 $\bar{K}N - \pi\Sigma$ interaction.}
\label{hyo4}
\end{center} 
\end{table*}

\begin{table*}[htbp]
\begin{center}
\begin{tabular}{c||ccc|c}
   &$\bar{K}N$(MeV)&$\pi\Sigma$(MeV)& $\pi\Lambda$(MeV)&
Scattering Length(fm) \\ \hline \hline
(A) & 1100 & 850 & 1250 & $0.68+i0.60$ \\
\hline
(B) & 950  & 800 & 1250 & $0.65+i0.46$ \\ 
\hline
\end{tabular}
\caption{The cutoff parameters, scattering length of the
relativistic models of
I=1 $\bar{K}N - \pi Y$ interaction.}
\label{hyo5}
\end{center} 
\end{table*}

\begin{table*}[htbp]
\begin{center}
\begin{tabular}{c||ccc|c}
   &$\bar{K}N$(MeV)&$\pi\Sigma$(MeV)& $\pi\Lambda$(MeV)&
Scattering Length(fm) \\ \hline \hline
(A) & 920 & 960 & 640 & $0.72+i0.59$ \\
\hline
(B) & 800 & 940 & 660 & $0.68+i0.45$ \\ 
\hline
\end{tabular}
\caption{The cutoff parameters, scattering length of the
nonrelativistic models of
I=1 $\bar{K}N - \pi Y$ interaction.}
\label{hyo6}
\end{center} 
\end{table*}

\begin{figure*}
\resizebox{1.0\textwidth}{!}{%
\includegraphics{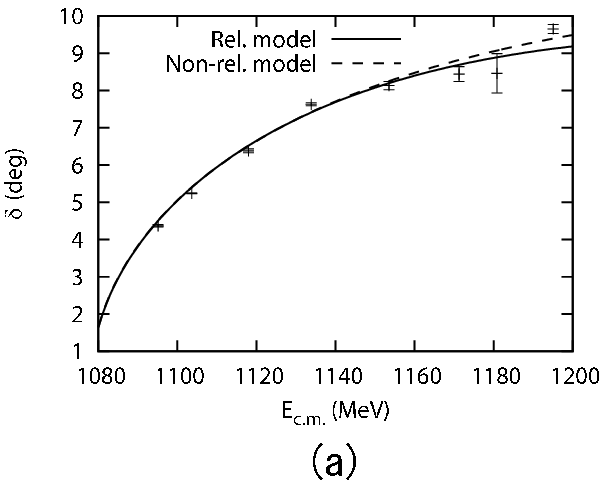}\hspace*{0.5cm}
\includegraphics{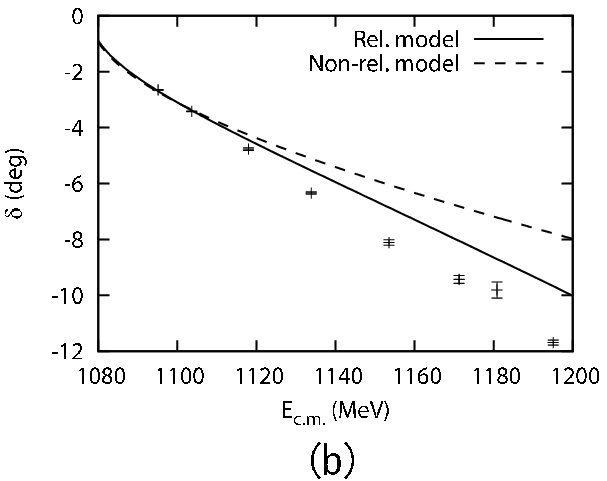}}
\caption{The phase shift of the $\pi N$ scattering for $S_{11}$ (a),
and $S_{31}$ (b) partial waves. The solid curve shows the relativistic model
and the dashed curve shows the nonrelativistic model.
Data are taken from Ref.~\cite{VPI}.}
\label{piN}
\end{figure*}

The form of the s-wave $\pi N$ interactions
is taken as  Eq. (\ref{mb-int}).
The constant $C_{\alpha ,\beta}$ is
$4$ for $I=1/2$ and $-2$ for $I=3/2$ states.
The parameters of the potentials are determined by fitting
the scattering length and the low energy phase shifts.
For $I=1/2$ state, the strength of the potential
is modified as $\lambda C_{\beta,\alpha}$
by introducing a phenomenological factor $\lambda$
to describe the data of the scattering length
$(0.1788 \pm 0.0050)m_{\pi}^{-1}$~\cite{piNsl} and the
phase shifts~\cite{VPI}.
The fitted parameters $\lambda$ and $\Lambda$ are
shown in Table \ref{hyo7} together with the scattering
length calculated using the models.
The models describe well the $S_{11}$ phase shifts
up to $1.2$ GeV as shown in Fig. \ref{piN}.

\begin{table*}[htbp]
\begin{center}
\begin{tabular}{c||cc|c}
   & $\lambda$ & $\Lambda$(MeV)& scattering length \\ \hline \hline
Relativistic   & 0.90 & 800 & $0.175 m_{\pi}^{-1}$ \\
\hline
Nonrelativistic & 0.85 & 800 & $0.177 m_{\pi}^{-1}$ \\ 
\hline
\end{tabular}
\caption{Parameters and scattering length for 
the relativistic and nonrelativistic model of
I=1/2 $\pi N$ interaction.}
\label{hyo7}
\end{center} 
\end{table*}

For the $I=3/2$ $\pi N$ scattering,
the $\pi N$ potential is constructed so as to reproduce 
the scattering length 
$(-0.0927 \pm 0.0093)m_{\pi}^{-1}$~\cite{piNsl} and 
the $S_{31}$ partial wave phase shifts data.
Here we introduced a modified dipole form factor as
\begin{eqnarray}
g(\vec p)=\frac{\Lambda ^4}{(\vec p ^2 + \Lambda ^2)^2}
\times (1+a \vec p ^2).
\end{eqnarray}
The parameters of the model are $\Lambda$ and $a$ for the form factor
and the strength parameter $\lambda$.
The obtained parameters are summarized in Table \ref{hyo8}.
The relativistic model can describe well the 
phase shifts up to $1.2$ GeV as shown in Fig. \ref{piN}; however,
the non-relativistic model starts to deviate from the data
at around $1.1$ GeV.

\begin{table*}[htbp]
\begin{center}
\begin{tabular}{c||ccc|c}
   & $\lambda$ & $\Lambda(MeV)$& $a(fm)^2$ & scattering length
 \\ \hline \hline
Relativistic   & 2.7 & 618 & 0.50 & $-0.095 m_{\pi}^{-1}$ \\
\hline
Nonrelativistic & 3.0 & 628 & 0.30 & $-0.101 m_{\pi}^{-1}$ \\ 
\hline
\end{tabular}
\caption{Parameters and scattering length for 
the relativistic and nonrelativistic model of
I=3/2 $\pi N$ interaction.}
\label{hyo8}
\end{center} 
\end{table*}

\begin{figure*}
\resizebox{0.45\textwidth}{!}{%
\includegraphics{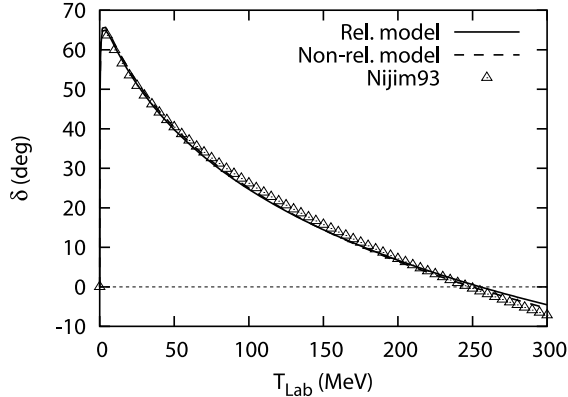}}
\caption{Phase shifts of the $NN$ scattering for $^1S_0$ state. 
Solid curve shows the relativistic model
and dashed curve shows the nonrelativistic model.
The phase shifts calculated from the model of Ref.~\cite{NNps}
are shown in triangles.
}
\label{NN}
\end{figure*}

We used a Yamaguchi-type separable interaction 
for the nucleon-nucleon potential.
To take into account the long range attractive interaction and
the short range repulsion of the two-nucleon interaction,
we used a two-term separable potential,
\begin{eqnarray}
<\vec{p}'|V_{BB}|\vec{p}>
  =  C_R g_R(\vec{p}') g_R(\vec{p})
- C_A g_A(\vec{p}') g_A(\vec{p}).
\label{NN-int}
\end{eqnarray}
Here $C_R$ ($C_A$) is the coupling strength 
of the repulsive (attractive) potential.
$g_R(\vec p)$ ($g_A(\vec p)$) is the form factor,
whose form is given as
$g_R(\vec p) = \Lambda_R ^2 /(\vec p ^2 + \Lambda_R ^2)$
($g_A(\vec p) = \Lambda_A ^2 /(\vec p ^2 + \Lambda_A ^2)$),
where $\Lambda$ is a cutoff of the nucleon-nucleon potential.
The adjustable parameters in our nucleon-nucleon potential are determined
by fits to the data of the $^1 S_0$ phase shifts~\cite{NNps}.
The best-fit parameters are summarized in Table \ref{hyo8}.
The low energy phase shifts of the $^1S_0$ state is shown in Fig. \ref{NN}.

\begin{table*}[htbp]
\begin{center}
\begin{tabular}{c||cccc}
   & $\Lambda _R(MeV)$ & $\Lambda _A(MeV)$& $C_R(MeV fm^3)$ &
$C_A(MeV fm^3)$ \\ \hline \hline
Relativistic   & 1144 & 333 & 5.33 & 5.61 \\
\hline
Nonrelativistic & 1215 & 352 & 5.05 & 5.84 \\ 
\hline
\end{tabular}
\caption{Our parameters of 
the relativistic and nonrelativistic model for
$NN$ scattering.}
\label{hyo9}
\end{center} 
\end{table*}

\section{Results and Discussion}

The dibaryon resonance with  $J^\pi=0^-$, $S=-1$, $I=1/2$
is studied using a formalism of the Faddeev equation as explained in section 2.
We assume all the angular momentum to be in an s-wave state
and the spin singlet state $S_{BB}= 0$ for the two baryon states.
We have included the dominant 
$\bar{K}NN$, $\pi\Sigma N$ and $\pi\Lambda N$ Fock-space components,
whose
isospin wave functions are
$[\bar{K}\otimes [NN]_{I=1}]_{I=1/2}$,
$[\pi\otimes [\Sigma N]_{I=1/2,3/2}]_{I=1/2}$ and 
$[\pi\otimes [\Lambda N]_{I=1/2}]_{I=1/2}$.
An approximation within this model is that 
the weak $Y N$ interaction is not included.

\begin{figure}
\resizebox{0.45\textwidth}{!}{%
\includegraphics{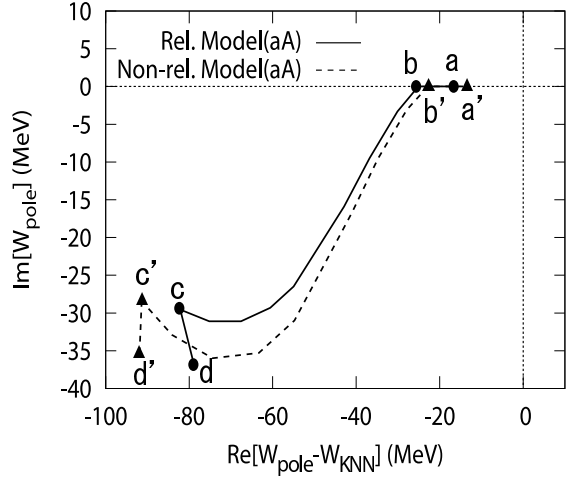}}
\caption{The pole trajectories of the $\bar KNN-\pi Y N$ scattering
amplitude for the $J^{\pi}=0^-$ and $I=1/2$ state.
The solid curve and the filled circles (dashed curve and filled
 triangles) show the
results of the relativistic (nonrelativistic) model (aA).
Here $W_{KNN} = m_K + 2 m_N$.
}
\label{Fig-traject}
\end{figure}

Let us start to 
 examine the three-body resonance energy
by taking into account only the $\bar{K}N$ interactions 
$v_{\bar{K}N-\bar{K}N}^{I=0,1}$ neglecting the $\pi Y N$ Fock space.
In this case, 
the bound state pole is expected to lie on the physical Riemann sheet
below $m_K + 2 m_N$ 
if the $\bar{K}N$ attraction is strong enough.
Therefore it is not necessary to use the
analytic continuation of the amplitude
with the deformed contour discussed in section 2,
so we simply use the integral over the momentum $p_i$ in the real axis.
The results are shown in Fig. \ref{Fig-traject}
marked by $a$ and $a'$ for the the 'relativistic' and 
'non-relativistic' models.
Here we use the 'standard' parameters (aA) of the $\bar{K}N$ interaction 
with nonrelativistic and relativistic kinematics.
The binding energies are about 18 MeV. 
The $\bar{K}N$ interaction included in $\tau$ and $Z$
is strong enough to bind the $\bar{K}NN$ system,
where the $I=0$ $\bar{K}N$ interaction plays a dominant role.
We then take into account the $NN$ interaction. Then the binding energy increased furthermore
 to 25.1 MeV (22.8 MeV) shown as $b$ ($b'$) for 'relativistic'
('nonrelativistic') model.
Notice that if we neglect the repulsive component of the
$NN$ interaction, we obtain a much more deeply bound state.

\begin{figure*}
\resizebox{0.45\textwidth}{!}{%
\includegraphics{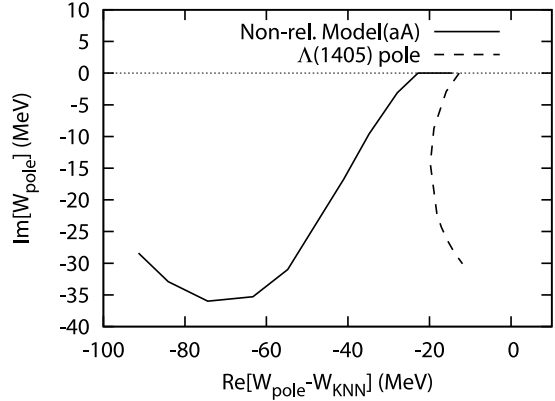}
}
\caption{
The pole trajectories of
the three-body resonance (solid curve) and $\Lambda(1405)$ (dashed
 curve) using the nonrelativistic model of (aA).
}
\label{sub-cut}
\end{figure*}

In the next step, we gradually include the $\pi Y N$ interactions,
while the pion-exchange $Z$ diagram is not yet included.
To do this, we multiply by factor $x$ 
the coupling constants $C_{\alpha,\beta}$ of the $\bar{K}N-\pi Y$
and $\pi Y-\pi Y$ interactions as $x C_{\alpha,\beta}$.
When the parameter is zero, $x=0$, the $\pi Y$ is disconnected from
$\bar{K}N$ and when it takes the value 1, $x=1$, we recover the full model.
By varying the parameter $x$ from $0$ to $1$, we can
follow the trajectory of the resonance pole from the bound state pole.
Now the $\bar{K}NN$ bound state decays into the $\pi Y N$ channel
and the bound state pole moves into the unphysical sheet.
Since the $\bar{K}NN$ bound state was found above the $\pi\Sigma N$ threshold,
the resonance pole may be on the $\pi Y N$ unphysical and
$\bar{K}NN$ physical Riemann sheet, which we have discussed in section 2. 
The results of the pole trajectories 
are shown by the solid and dashed curves in Fig. \ref{Fig-traject}
corresponding to the relativistic and the nonrelativistic models. 
Increasing the coupling to the $\pi Y N$ channel causes the width
as well as the binding energy of the resonance increases.
For larger binding energy $Re(W_{pole}-W_{KNN}) < -60$ MeV the width
starts to decrease because of the decreasing 
phase space for the decay into the $\pi \Sigma N$ state. 
The pole position is at $-82 - i29$ MeV ($-91 - 28i$)
for the relativistic (nonrelativistic) model shown as $c$($c'$).
It is noticed that the numerical method to follow the pole trajectories
helps us to find whether we encounter singularities or not.
As an example, the pole of the three-body resonance is shown by the solid
line in Fig. \ref{sub-cut} for $0< x < 1$.
The pole of the $\Lambda(1405)$ is also shown by the dashed curve.
The trajectory of the $\bar{K}NN$ resonance is similar to the 
the case A in Fig. \ref{lam-cut}, and the integration contour does not
intercept the singularity arising from the two-body resonance $\Lambda(1405)$.
%

Finally we include the
$\pi$ exchange mechanism in $Z$ 
and $\pi N$ two-body scattering terms in $\tau$,
which adds another mechanism for
the decay of the $\bar{K}NN$ into $\pi Y N$
and is important for the width of the three-body resonance.
The final results of the $\bar KNN-\pi Y N$ resonance poles 
are denoted by $d$ and $d'$ in Fig. \ref{Fig-traject}.
This mechanism increases the width of the three-body resonance
by about 14 MeV, while the effect on the real part is small.
The cancellation between
the attractive $I=1/2$ $\pi N$ interaction and the
repulsive $I=3/2$ $\pi N$ interaction may lead to the
small effects on the real part of the resonance energy.
The effects of $\pi \Lambda N$ channel is small which increases
binding energy and half-width at most by 1MeV. 
 The pole position of the three-body resonance
is $W = M - i \Gamma/2 = 2m_N+m_K - 79.3 - i37.1$ MeV 
($2m_N + m_K - 92.2 - i35.4$ MeV)
for relativistic (nonrelativistic) model shown as $d$ and $d'$.


\begin{table*}[htbp]
\begin{center}
\begin{tabular}{c||c|c}
    & Model (A)     & Model (B) \\
\hline \hline
(a) & $-79.3-i37.1$  & $-79.3-i37.3$ \\
\hline
(b) & $-93.3-i27.4$  & $-93.3-i27.6$ \\
\hline
(c) & $-57.2-i38.6$  & $-56.9-i38.6$ \\
\hline
(d) & $-72.4-i31.7$  & $-72.2-i31.9$ \\
\hline
(e) & $-87.1-i40.8$  & $-87.1-i41.0$ \\
\hline
(f) & $-63.3-i22.2$  & $-63.2-i22.3$ \\
\hline
\end{tabular}
\caption{The pole energy ($W_{pole} - m_K - 2 m_N$)
of the three-body resonance using relativistic models.
The listed pole energies in MeV
can be related to binding energy $B$ and the width $\Gamma$
as  $W_{pole} - m_K - 2 m_N = - B-i \Gamma /2$.
}
\label{hyo10}
\end{center} 
\end{table*}
\begin{table*}[htbp]
\begin{center}
\begin{tabular}{c||c|c}
    & Model (A)     & Model (B)    \\
\hline \hline
(a) & $-92.2-i35.4$  & $-92.3-i35.6$ \\
\hline
(b) & $-101.6-i20.7$ & $-101.6-i20.7$ \\
\hline
(c) & $-72.7-i53.9$  & $-72.5-i54.9$ \\
\hline
(d) & $-83.0-i33.3$  & $-83.0-i33.6$ \\
\hline
(e) & $-98.1-i33.2$  & $-98.2-i33.3$ \\
\hline
(f) & $-66.5-i24.4$  & $-66.3-i24.4$ \\
\hline
\end{tabular}
\caption{The pole energy of the three-body resonance.
The same as Table \ref{hyo10} but for the
nonrelativistic models.}
\label{hyo11}
\end{center} 
\end{table*}


The model dependence of our results on the three-body resonance
is summarized in Tables \ref{hyo10} and \ref{hyo11}.
The $\bar KNN - \pi Y N$ resonance pole is located on
the $\bar KNN$ physical and $\pi Y N$ unphysical sheet
with the binding energy, $B \sim 60-95$ MeV,
and the width, $\Gamma \sim 45-80$ MeV, using relativistic models.
All of our models predict resonance energies above the
$\pi\Sigma N$ threshold.
The relatively large model dependence of our results
is due to the uncertainty in the
models of $I=0$ $\bar KN - \pi \Sigma$ interaction.
Comparing the results of model (A) with model (B),
we can see the three-body pole position is almost independent of
the parameters of the $I=1$ $\bar KN - \pi Y$ interaction.
By varying the real(imaginary) part of the fitted scattering length by
$\pm 0.1$ fm,
the binding energy of the three-body resonance is affected by
$\sim \pm 14 (8) $ MeV.
Following another way to construct the model, the parameters of the model (f)
are fitted to the pole energy of the $\Lambda (1405)$.
This model predicts the scattering length $-1.72+i0.44$ fm.
The energy of the three-body resonance is found to be
$B =63$ MeV with a rather small width, $\Gamma = 44$ MeV,
compared with the models (a-e), which can already be seen in the
small imaginary part of the scattering length in model (f).

Let us briefly compare our results with those of the other theoretical studies
of the $K^- pp$ resonance, which use a nonrelativistic approach.
Our resonance has a deeper binding energy and a similar
width compared with those in Ref.~\cite{Aka2}.
However, it is not straightforward to compare with the pole energy
of Ref.~\cite{Aka2} because of the differences in the
method to obtain the three-body resonance energy and on the
model for the $\bar{K}N$ interaction. Their $\bar{K}N$ potential
is stronger and has a short range than ours.
Recently Shevchenko, Gal and Mares~\cite{she} studied $K^-pp$ system using
the nonrelativistic coupled-channel Faddeev equation.
Though the details of their method is not described in Ref.~\cite{she},
it seems their approach is quite similar to our present study. 
They employed a phenomenological $\bar{K}N$ potential model,
and reported $B \sim 55 - 70$ MeV and $\Gamma \sim  95 - 110$ MeV. 
Their result is consistent with our results of the nonrelativistic model.
Specially our result using the model (c) gives a quite similar 
resonance energy and width.

In summary we have studied the existence and properties of
a strange dibaryon resonance using the
$\bar{K}NN-\pi Y N$ coupled channel Faddeev equation.
By solving the three-body equation the energy dependence of
the resonant $\bar{K}N$ amplitude is fully taken into account.
The resonance pole has been investigated 
from the eigenvalue of the kernel
with the analytic continuation of the  scattering amplitude
on the unphysical Riemann sheet.
The model of the $\bar{K}N-\pi Y$ interaction is 
constructed from the leading order term of the chiral Lagrangian
takes into account the relativistic kinematics.
The $\bar{K}N$ interaction parameters are
fitted to the scattering length given by Martin.
We found a resonance pole 
at $B \sim 79$ MeV and $\Gamma \sim 74$ MeV
in the relativistic model (aA). 
However, as the $\bar{K}N$ interaction is not very well constrained by the data, 
we studied a possible range of the resonance energies 
by considering different parameter sets of the $\bar KN - \pi Y$ interaction.
The binding energy and the full width can be
in the range of   $B \sim 60 - 95$ MeV and 
$\Gamma \sim  45 - 80$ MeV when computed in the relativistic model.
In order to connect the resonance found in this work to the experimental signal,
further theoretical studies on the production mechanism and further decay of the
resonance especially to the $\Lambda-p$ channel are necessary.

\begin{acknowledgments}
The authors are grateful to Prof. A. Matsuyama for very useful
discussions on the three-body resonance.
We also thank Drs. B. Juli\'a-D\'{\i}az, T.-S. H. Lee and Prof. A. Gal 
for discussions.
This work is supported
by a Grant-in-Aid for Scientific Research on Priority Areas(MEXT), Japan
with No. 18042003. 
\end{acknowledgments}
\appendix*
\section{}

The spin-isospin recoupling coefficient of the particle
exchange interaction $Z$ is briefly explained.
The coefficient  given in Eq. (1.181) of Ref. \cite{book}
can be simplified for s-wave states.
The three-body state with total spin and isospin
$(S_{tot},I_{tot})$, which couples with baryon 'isobar' with spin
and isospin $(S,I)$ and the spectator baryon $B_{i(S_i,I_i)}$,
is given as
\begin{eqnarray}
|[[M_{3(S_3,I_3)} \otimes B_{j(S_j,I_j)}]_{(S,I)} 
\otimes B_{i(S_i,I_i)}]_{(S_{tot},I_{tot})}>.
\end{eqnarray}
Here baryon  $i,j$ represents particle 1 or 2, and the meson is always
assigned as the third particle.
The wave function of the three-body state, which couples
with  dibaryon 'isobar' and the spectator meson $M_3$, is given as 
\begin{eqnarray}
|[[B_{1(S_1,I_1)} \otimes B_{2(S_2,I_2)}]_{(S,I)}
 \otimes M_{3(S_3,I_3)}]_{(S_{tot},I_{tot})}>.
\end{eqnarray}
Then  Eq. (\ref{Z}) is extended to include
spin-isospin degrees of freedom.
The particle exchange interaction 
for the spectators $l,m$, the isobars $f',f$ with spin-isospin
$(S',I')$ and $(S,I)$ and the exchanged particle $n$
can be expressed as follows
\begin{eqnarray}
Z_{l,f'(S',I'), m,f(S,I)}(p_l, p_m, W) & = & 
R_{l,f'(S',I'), m,f(S,I)}  
\int d(\hat{p_l} \cdot \hat{p_m}) \frac{
2\pi g_{f'(S',I')}(q_l) g_{f(S,I)}(q_m)}
{W-E_l(p_l)-E_m(p_m)-E_n(\vec p_l + \vec p_m)}, \nonumber \\
\label{Z-full}
\end{eqnarray}
where $f$ represents isobar $Y_K,Y_\pi,d$ and $N^*$.

$R_{l,f'(S',I'),m,f(S,I)}$ is given by 
the overlap of the initial and final spin-isospin wave functions.
For the meson$(M_3)$ exchange mechanism,
 $R_{i,f'(S',I'),j,f(S,I)}$ is given as,
\begin{eqnarray}
R_{i,f'(S',I'),j,f(S,I)} &=&
<[[M_{3(S_3,I_3)} \otimes B_{j(S_j,I_j)}]_{(S',I')} 
\otimes B_{i(S_i,I_i)}]_{(S_{tot},I_{tot})}| \nonumber \\
&&\times |[[M_{3(S_3,I_3)} \otimes B_{i(S_i,I_i)}]_{(S,I)} 
\otimes B_{j(S_j,I_j)}]_{(S_{tot},I_{tot})}> \nonumber \\
&=&
 (-1)^{S+S'-S_3-S_{tot}}W(S_i,S_3,S_{tot},S_j;S,S')\sqrt{(2S+1)(2S'+1)} 
 \nonumber \\
& & \times (-1)^{I+I'-I_3-I_{tot}}W(I_i,I_3,I_{tot},I_j;I,I')
\sqrt{(2I+1)(2I'+1)}.
\end{eqnarray}
For the baryon($B_j$) exchange mechanism, 
$R_{i,f'(S',I'),3,f(S,I)}$ is given as,
\begin{eqnarray}
R_{i,f'(S',I'),3,f(S,I)} &=&
<[[M_{3(S_3,I_3)} \otimes B_{j(S_j,I_j)}]_{(S',I')} 
\otimes B_{i(S_i,I_i)}]_{(S_{tot},I_{tot})}|\nonumber \\
&&\times |[[B_{1(S_1,I_1)} \otimes B_{2(S_2,I_2)}]_{(S,I)} 
\otimes M_{3(S_3,I_3)}]_{(S_{tot},I_{tot})}> \nonumber \\
&=& (-1)^{S_3+S-S_{tot}+I_3+I-I_{tot}}
W(S_3,S_j,S_{tot},S_i;S',S)\sqrt{(2S'+1)(2S+1)} \nonumber \\
&&\times W(I_3,I_j,I_{tot},I_i;I',I)\sqrt{(2I'+1)(2I+1)} \nonumber \\
&&\times 
(\delta _{i,2}\delta _{j,1} + \delta _{i,1}\delta _{j,2} 
(-1)^{S_i+S_j-S+I_i+I_j-I}). \label{eqap}
\end{eqnarray}

When we anti-symmetrize the AGS equation, the last factor
in the bracket in Eq. (\ref{eqap}) projects the anti-symmetric
two nucleon states. This can be explicitly seen by comparing the 
exchange of nucleon 2 $Z_{2,Y_{K}(S',I'),3,d(S,I)}$ and nucleon 1
$Z_{1,Y_{K}(S',I'),3,d(S,I)}$ interactions.
Using Eq. (\ref{eqap}), those interactions are related as
\begin{eqnarray}
 R_{1,Y_{K}(S',I'),3,d(S,I)} &=& (-1)^{S+I} R_{2,Y_{K}(S',I'),3,d(S,I)},
\end{eqnarray}
which leads to  Eq. (\ref{ags-full}) as
\begin{eqnarray}
X_{Y_K,Y_K} = (1 - (-1)^{S+I})Z_{Y_K,d}\tau_{d,d}X_{d,Y_K} + \cdots .
\end{eqnarray}



\begin{thebibliography}{99}

%
\bibitem{Gal1}J. Mares, E. Friedman and A. Gal, Nucl. Phys. \textbf{A770}, 84 (2006).
\bibitem{Ram}L. Tol\'os, A. Ramos and E. Oset, Phys. Rev. C \textbf{74}, 015203 (2006).
%
%
\bibitem{Aka1}Y. Akaishi and T. Yamazaki, Phys. Rev. C \textbf{65}, 044005 (2002).
\bibitem{Aka2}T. Yamazaki and Y. Akaishi, Phys. Lett. \textbf{B535}, 70 (2002).
\bibitem{Dote}A. Dote, H. Horiuchi, Y. Akaishi and T. Yamazaki, Phys. Rev. C \textbf{70}, 044313 (2004).
%
%
\bibitem{Agne}M. Agnello et al., Phys. Rev. Lett. \textbf{94}, 212303 (2005).
%
%
\bibitem{Magas}V.K. Magas, E. Oset, A. Ramos and H. Toki, Phys. Rev. C \textbf{74}, 025206 (2006).
%
\bibitem{Glock}W. Gl\"ockle, Phys. Rev. C \textbf{18}, 564 (1978). 
\bibitem{Moll}K. M\"oller, Czech. J. Phys. \textbf{32}, 291 (1982).
%
\bibitem{Matsu}A. Matsuyama and K. Yazaki, Nucl. Phys. \textbf{A534}, 620 (1991). \\
A. Matsuyama, Phys. Lett. \textbf{B408}, 25 (1997).
\bibitem{Afnan2}I.R. Afnan and A.W. Thomas, Phys. Rev. C \textbf{10}, 109 (1974).
%
\bibitem{Pear}B.C. Pearce and I.R. Afnan, Phys. Rev, C \textbf{30}, 2022 (1984).
\bibitem{Afnan}I.R. Afnan and B.F. Gibson, Phys. Rev. C \textbf{47}, 1000 (1993).
%
%
\bibitem{Oller}J.A. Oller and U.-G. Mei\ss ner, Phys. Lett. \textbf{B500}, 263 (2001).
\bibitem{Jido}D. Jido et al., Nucl. Phys. \textbf{A725}, 181 (2003).
\bibitem{Bor}B. Borasoy, R. Ni\ss ler and W. Weise, Eur. Phys. J. A \textbf{25}, 79 (2005).
\bibitem{Hama}T. Hamaie, M. Arima and K. Masutani, Nucl. Phys. \textbf{A591}, 675 (1995).
%
%
\bibitem{ikeda}Y. Ikeda and T. Sato, arXive:nucl-th/0701001.
\bibitem{she}N.V. Shevchenko, A. Gal and J. Mares, Phys. Rev. Lett. \textbf{98}, 082301 (2007).
%
%
\bibitem{ags}E.O. Alt, P. Grassberger and W. Sandhas, Nucl. Phys. \textbf{B2}, 167 (1967).
%
\bibitem{book} I.R. Afnan and A.W. Thomas, \textit{in Modern Three-Hadron
Physics}, edited by A.W. Thomas (Springer, Berlin, 1977), Chap. 1.
%
%
%
\bibitem{Mart}A.D. Martin, Nucl. Phys. \textbf{B179}, 33 (1981).
\bibitem{Bora2}B. Borasoy, U.-G. Mei\ss ner and R. Ni\ss ler, Phys. Rev. C \textbf{74}, 055201 (2006).
%
\bibitem{Dlitz}R.H. Dalitz and A. Deloff, J. Phys. G \textbf{17}, 289 (1991).
%
\bibitem{Iwa}M. Iwasaki et al., Phys. Rev. Lett. \textbf{78}, 3067 (1997).
\bibitem{Itoh}T.M. Ito et al., Phys. Rev. C \textbf{58}, 2366 (1998).
\bibitem{Dear}G. Beer et al., Phys. Rev. Lett. \textbf{94}, 212302 (2005).
%
\bibitem{bahaoui}A. Bahaoui, C. Fayard, T. Mizutani and B. Saghai,
            Phys. Rev. C \textbf{68}, 064001 (2003).
%
\bibitem{crs1}W.E. Humphrey and R.R. Ross, Phys. Rev, \textbf{127}, 1305 (1962).
\bibitem{crs2}M. Sakitt et al., Phys. Rev. \textbf{139}, B719 (1965).
\bibitem{crs3}J.K. Kim, Phys. Rev. Lett. \textbf{14}, 29 (1965).
\bibitem{crs4}W. Kittel, G. Otter and I. Wacek, Phys. Lett. \textbf{B21}, 349 (1966).
\bibitem{crs5}D. Evans et al., J. Phys. G \textbf{9}, 885 (1983).
%
\bibitem{piNsl}H.-Ch. Schr\"oder et al., Phys. Lett. \textbf{B469}, 25 (1999).
%
\bibitem{VPI}
R.A. Arndt, I.I. Strakovsky, R.L. Workman and M.M. Pavan,
      Phys. Rev. C \textbf{52}, 2120 (1995).\\
R.A. Arndt, I.I. Strakovsky and R.L. Workman,
      Int. J. Mod. Phys. A \textbf{18}, 449 (2003).
%
%
\bibitem{NNps}V.G.J. Stoks, R.A.M. Klomp, C.P.F. Terheggen and J.J. de Swart, Phys. Rev. C \textbf{49}, 2950 (1994).
%
\end{thebibliography}
\end{document}